\RequirePackage{fix-cm}
\documentclass[smallextended]{svjour3}       
\smartqed  
\usepackage{graphicx}
\usepackage{comment}

\begin{document}

\title{Determination of the polarization observables $C_{x}$, $C_{z}$, and $P$ for the $\vec{\gamma}d\to K^{0}\Lambda(p)$ reaction \thanks{
Work supported in part by the U.S. NSF under grant PHY-1505615.
}
}


\author{Colin Gleason \and Yordanka Ilieva \and on behalf of the CLAS Collaboration 
}


\institute{C. Gleason \at
              Indiana University \\
              727 E. Third St., Bloomington, IN 47405 \\
              \email{gleasonc@jlab.org}             \\
           \and
           Y. Ilieva \at
              University of South Carolina\\
              712 Main St., Columbia, SC 29208
}

\date{Received: date / Accepted: date}

\maketitle

\begin{abstract}
Many excited nucleon states ($N^{*}$s) predicted by quark models, but not observed in $\pi N$ channels, are expected to couple strongly to kaon--hyperon ($KY$) channels. 
While in the last decade data have been published for $KY$ photoproduction off the proton, data off the neutron are scarce.
In this paper we present preliminary results for the polarization observables $P$, $C_x$, and $C_z$ for the reaction $\gamma d\to K^{0}\Lambda(p)$, where $(p)$ denotes
the spectator proton.
 The observables cover photon energies, $E_{\gamma}$, between 0.9 GeV and 2.6 GeV and kaon center-of-mass angles, cos$\theta_{K^{0}}^{CM}$, between $-0.9$ and 1. 
The data were collected in experiment E06-103 (g13) with the CLAS detector at the Thomas Jefferson National Accelerator Facility using a circularly-polarized photon beam and an unpolarized liquid deuterium target. 
We also discuss the effect of neutron binding on the observables. 
Our study is part of a broader effort by the g13 group to provide cross sections and polarization observables for meson photoproduction off the neutron and is expected to have a significant impact on the $N^{*}$ research.
\end{abstract}

\section{Introduction}
\label{intro}
Mapping the excited nucleon ($N^{*}$) spectrum plays a crucial role in understanding the underlying degrees of freedom in Quantum Chromodynamics (QCD).
Recently, there has been significant interest in the study of strangeness photoproduction, specifically with $K\Lambda$ and $K\Sigma$ final states.
This is because many $N^{*}$ states above 1800 MeV are predicted to have significant branching ratios into strangeness channels while the $\pi N$ photocoupling is significantly smaller \cite{PhysRevD.58.074011}, \cite{PhysRevD.94.074040}.
The majority of strangeness photoproduction data come from free proton targets (see References \cite{Bradford:2006ba}, \cite{McNabb:2003nf}, and \cite{PhysRevC.88.035209} for recent examples). 
Since $N^{*}$ states have different photocouplings to protons and neutrons, it is expected that experimental observables, such as cross sections and polarization observables, from neutron targets will provide new information about poorly established $N^{*}$ states above 1800 MeV.

Recently, the CLAS Collaboration has published the differential cross section of the $\gamma d\to K^{0}\Lambda(p)$ reaction~\cite{Compton:2017xkt}.
When these cross sections are included in the data fitted by the partial-wave analysis (PWA) of the Bonn--Gatchina group, two possible solutions are found~\cite{Compton:2017xkt}.
Both solutions describe the cross-section data for $K^{0}\Lambda$, and also for $K^{+}\Sigma^{-}$, production reasonably well. 
Data for the polarization observables of the reaction $\gamma d\to K^{0}\Lambda(p)$ should be able to resolve the ambiguity between the two solutions, or will lead to a new solution.

\begin{figure}
	\centering
  	\graphicspath{ {./} }
   	\includegraphics[width=0.7\textwidth]{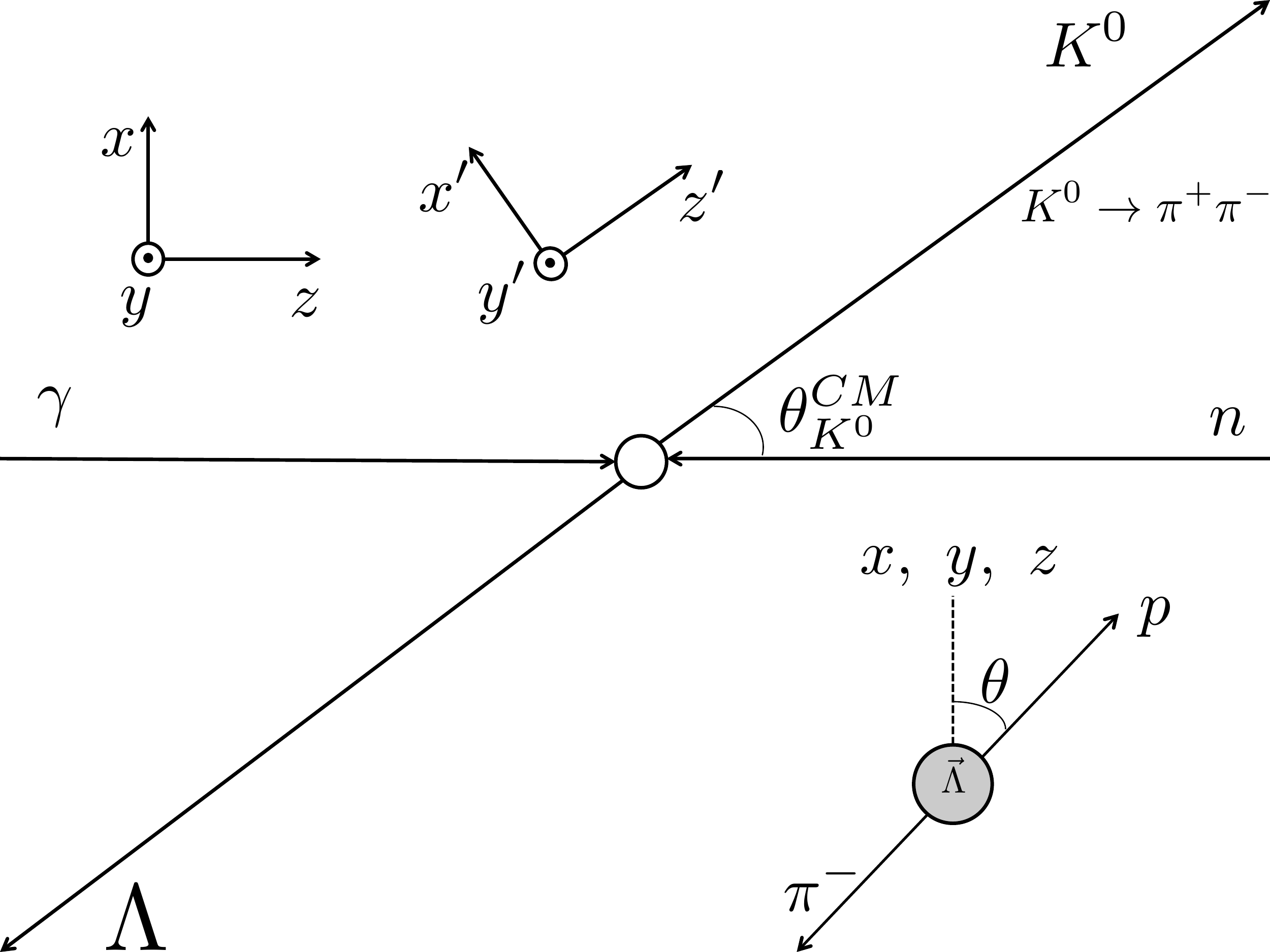} 
	\caption{A schematic diagram of the reference frame and the choice of coordinate axes used in this work. We consider the elementary process $\gamma n\to K^{0}\Lambda$., where the neutron is bound in the deuteron. The proton angle, $\theta$, is determined in the $\Lambda$ rest frame as the angle between the proton 3-vector and a coordinate axis. In the text, we use the labels $\theta_x$, $\theta_y$, and $\theta_z$ to denote the three possible angles.}
	\label{fig:reaction} 
\end{figure}

 The cross section for pseudoscalar meson photoproduction with circularly-polarized photon beam incident on unpolarized target can be expressed as 
 \begin{equation}
  \frac{d\sigma}{d\Omega}= \sigma_{0} [1-\alpha \cos \theta_{x}P_{circ}C_{x}+\alpha\cos\theta_{y}P-\alpha\cos\theta_{z}P_{circ}C_{z}],
 \end{equation}
where $\sigma_{0}$ is the unpolarized cross section, $P_{circ}$ is the degree of circular polarization of the photon beam, $\alpha=0.642$ is the self--analyzing power of $\Lambda$, $P$ is the induced polarization of the $\Lambda$, and $C_{x,z}$ are the double polarization observables that measure the polarization transfer from the photon to the $\Lambda$ \cite{NadelTuronski2008}.
Figure \ref{fig:reaction} shows a schematic diagram of the reaction $\gamma n\to K^{0}\Lambda$ and the definition of coordinate axes.
The incoming photon and the target nucleon interact to produce the final-state particles $K^{0}$ and $\Lambda$, which we reconstruct by detecting their decay products, $\pi^+\pi^-$ and $p\pi^-$, respectivley.
The angles $\theta_{x}$, $\theta_y$, and $\theta_{z}$ are the decay angles of the proton in the $\Lambda$ rest frame with respect to the $x$, $y$, and $z$ axis, respectively.
The objective of this work is to extract $C_{x}$, $C_{z}$, and $P$ for the reaction $\gamma d\to K^{0}\Lambda(p)$.

 \section{Experiment}
\label{sec:Experiment}
The data for this work were collected in Hall B of the Thomas Jefferson National Accelerator Facility (Jefferson Lab) as part of experiment E06-103 (g13) ~\cite{proposal}. 
The g13 run was divided into two run groups: g13a, which used a circularly-polarized photon beam, and g13b, which used a linearly-polarized photon beam.
Data for this analysis were taken during the g13a running where the circularly-polarized photon beam was incident upon a liquid deuterium target.
The photon beam was created in the Hall-B  photon tagging system \cite{Sober2000263} by the beam of electrons that was delivered by the Continuous Electron Beam Accelerator Facility (CEBAF)\cite{CEBAF2_paper}. 
Approximately 20 billion events were collected during the g13a run group.
The CEBAF Large Acceptance Spectrometer (CLAS) was used to detect multiple-charged-particle final states of nuclear reactions generated by the photon beam.
A complete description of CLAS is given in \cite{Mecking2003513}.
For the g13a experiment, the trigger required charged tracks in two of the six CLAS sectors resulting in an event rate of 10kHz with a dead time of $\approx$15\% \cite{NadelTuronski2008}.
This trigger condition was chosen to reduce accidental hits while enhancing final states with charged-particles.
%

\section{Reaction Selection}
The strange particles in the final state of the reaction $\gamma d\to K^{0}\Lambda (p)$, the $K^{0}$ and $\Lambda$, decay before they traverse through CLAS and, thus, cannot be measured directly.
We make use of the $K^{0}$ decay into a $\pi^{+}\pi^{-}$ pair and of the $\Lambda$ decay into a $p\pi^{-}$ to reconstruct the four-momentum vectors of the $K^{0}$ and the $\Lambda$.
These decay products, $\pi^+\pi^- p\pi^-$, are detected and identified using their speed and momentum as measured by CLAS. 
The photon that initiated the reaction was determined by means of the difference between the time when the photon arrives at the reaction vertex and the time when the fastest final-state particle detected in the CLAS was created at the vertex, known as coincidence time.
The fastest particle was chosen because none of the detected decay products originated from the event vertex.
CEBAF delivered electrons in bunches every 2.004 ns, while the trigger window for digitizing data by the
CLAS electronics was about 30 ns wide in the g13 experiment. This resulted in an average of about 20 reconstructed tagged photons per triggered event (event) that could have initiated the reaction.
An event was selected for further analysis if only one photon in that event was within a coincidence time window of $\pm$1 ns.

\begin{figure*}
	\graphicspath{ {./} }
   	\includegraphics[width=0.5\textwidth]{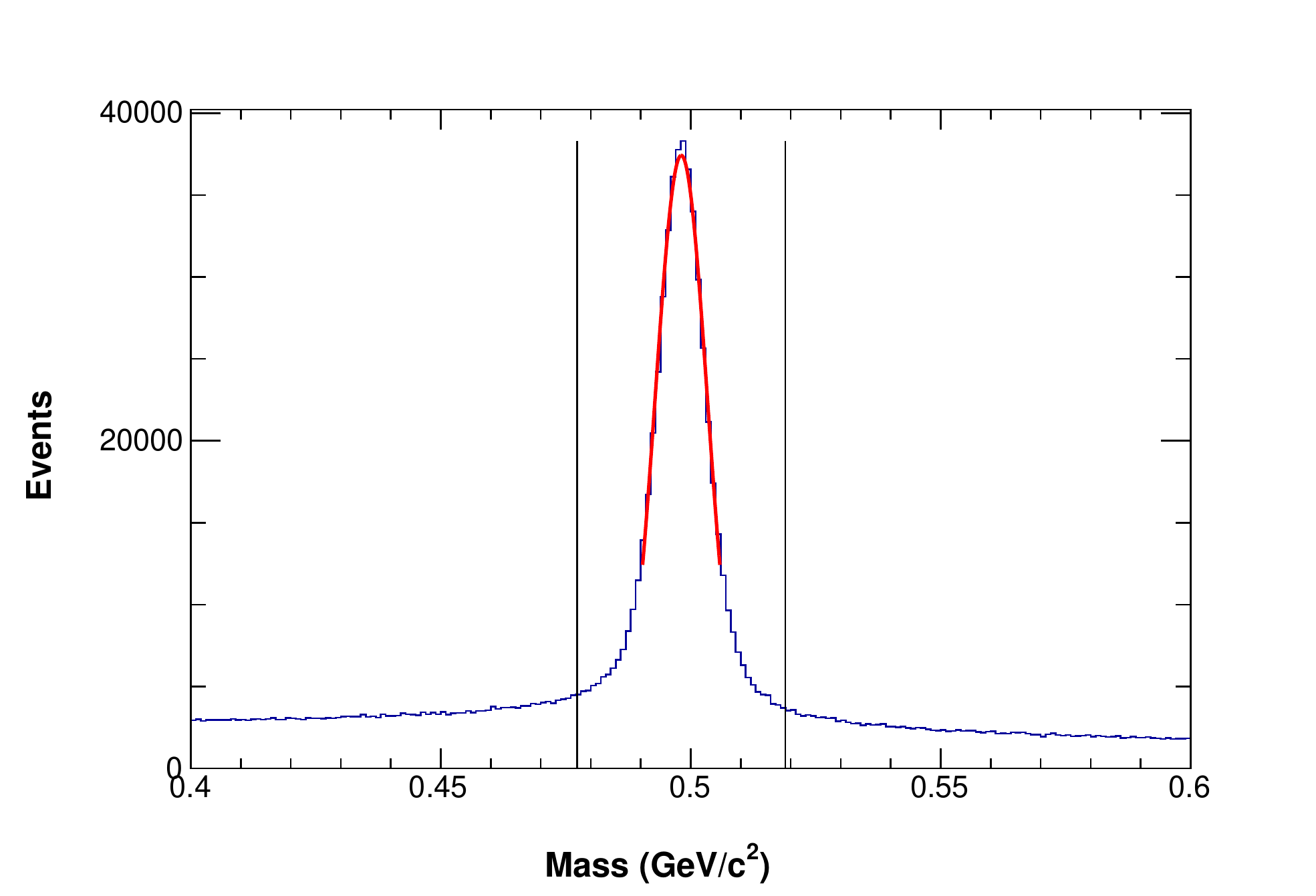} 
	\includegraphics[width=0.5\textwidth]{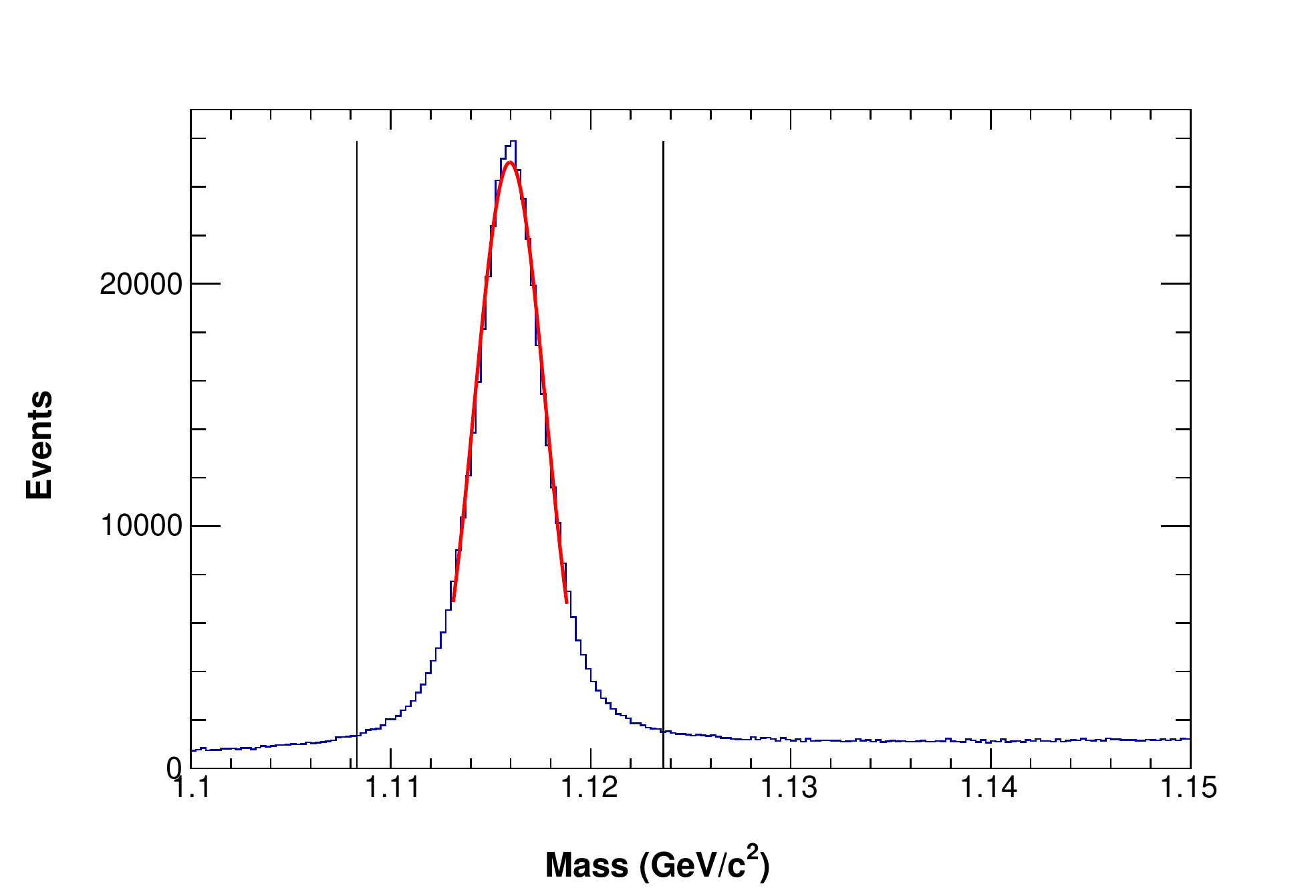} 
	\caption{Event distributions over the invariant mass of the $\pi^{+}\pi^{-}$ system (left) and the $p\pi^{-}$ system (right). One can clearly see the $K^{0}$ peak (left) and the $\Lambda$ peak (right). 
Each peak was fit 
with a Gaussian and the range of $\pm4\sigma$ from the mean was used to select events with a $K^{0}$ and a $\Lambda$.}
	\label{fig:InvariantMasses} 
\end{figure*}

The $K^{0}$ and $\Lambda$ were reconstructed by means of the invariant mass of the $\pi^{+}\pi^{-}$, $M(\pi^{+}\pi^{-})$, and $p\pi^{-}$, $M(p\pi^{-})$, respectively.
Figure \ref{fig:InvariantMasses} shows the event distributions over the invariant mass of the $K^{0}$ (left) and $\Lambda$ (right), $M(\pi^{+}\pi^{-})$ and $M(p\pi^{-})$, respectively.
Events within $\pm4\sigma$ of the mean of each fitted Gaussian were selected.
Due to the fact that the $K^0$ peak in the $M(\pi^{+}\pi^{-})$ distribution has long tails, this relatively broad selection allows us to keep for further analysis nearly all good $K^{0}$ candidates. 

The quasi-free reaction $\gamma n\to K^0\Lambda$, where $n$ is bound in the deuteron, is identified using the momentum of the final-state spectator proton.
Since it does not take part in the quasi-free process, the spectator proton in the final state is expected to have a momentum equal to its Fermi momentum. 
Thus, the momentum distribution of the final-state spectator protons in the reaction $\gamma d\to K^0 \Lambda (p)$ should be consistent with the Fermi momentum distribution of a nucleon bound in the deuteron. 
Studies performed with simulated data, where the spectator proton momentum is generated using the Paris potential, show that spectator momentum is indeed consistent with the Fermi momentum of a bound nucleon \cite{LACOMBE1981139}.
Since this proton is not detected, \textit{i.e.}, is ``missing," its momentum is referred to as ``missing" momentum. 
For each event, the missing momentum is reconstructed from the measured momenta using four-momentum conservation. 
We denote the magnitude of the missing 3-vector as $p_X$.
The quasi-free reaction was then identified by selecting events with $p_{X}<0.2$ GeV/$c$.
This cut eliminates most of the $K^{0}\Lambda p$ events produced in a more complex scattering, where the quasi-free mechanism was followed by final-state interactions, such as $\Lambda$ or $K^{0}$ re-scattering off the proton.

The final selection of good events makes use of the mass of the missing final state, $M_{X}$, in the reaction $\gamma d\to K^{0}\Lambda X$, where $M_{X}$ is defined as
\begin{equation}
	M_{X}=\sqrt{(\tilde{p}_{\gamma}+\tilde{p}_{d}-\tilde{p}_{K^{0}}-\tilde{p}_{\Lambda})^{2}},
\end{equation}
where $\tilde{p}_{\gamma},~\tilde{p}_{d},~\tilde{p}_{K^{0}},~\tilde{p}_{\Lambda}$ are the four momenta of the photon, deuteron target, $K^{0}$, and $\Lambda$, respectively. 
Only events within the mass range of 0.90--0.98 GeV/$c^{2}$ were kept for further analysis (see Section \ref{sec:BackgroundSub} for further discussion).

\section{Extraction of Observables}
\subsection{Background Subtraction} \label{sec:BackgroundSub}
The application of the selection criteria described in the previous section yields a reduced data sample (referred to as the \textit{total} sample).
This \textit{total} sample contains a large fraction of signal events and a smaller fraction of background events. 
The total sample was binned in ($W, ~\cos\theta_{K^{0}}^{CM}$). Here $W$ denotes the total energy in the center of mass of the $K^0\Lambda$ system and 
$\cos\theta_{K^{0}}^{CM}$ denotes the $K^{0}$ polar angle in the same system. 
Below we describe the procedure of extracting background-free observables from the total sample for each kinematic bin. The total sample contains events that
pass all the selection criteria described above.

There are two types of background events that remain in the sample.
The first type contains $p\pi^{+}\pi^{-}\pi^{-}$ events uncorrelated with $K^{0}\Lambda$ production. 
These events form a smooth background underneath the $K^{0}$ and the $\Lambda$ peaks in the invariant-mass distributions.
Our studies show that this background is not polarized.
The second type contains events from the photoproduction of higher-mass hyperons that decay into a final state containing a $\Lambda$.
An example is $K^{0}\Sigma^{0}$ production that is followed by a $\Sigma^{0}$ decay into a $\Lambda\gamma$. Here, the $\Lambda$ is indistinguishable from the $\Lambda$ in $K^{0}\Lambda$ production.
For each kinematic bin, the amount of these background events in the data sample is quantified by fits to the $M_X$ distribution.

The background-subtraction method used here is based on a 2016 CLAS Analysis Note for studying final-state interactions in the reaction $\vec{\gamma}d\to K^{+}\vec{\Lambda}n$ \cite{Zachariou_FSI_2016}.
For each kinematic bin, a set of \textit{total} observables, $P^{T},~C_{x}^{T}$, and $C_{z}^{T}$, is simultaneously extracted from the total data sample by means of the maximum log-likelihood method (see Sec. \ref{sec:ML_Method}).
Each \textit{total} observable is then corrected using signal-to-background ratios determined by fits to the $M_{X}$ distribution for each kinematic bin.
Simulation data (a Geant 3 simulation of CLAS~\cite{wolin_1996}) were used to determine the shape of the background in the $M_X$ distributions.
A complete discussion of the simulations and a derivation of the background-free observables $C_{x}$, $C_{z}$, and $P$ in the reaction $\gamma d\to K^{0}\Lambda (p)$ are given in \cite{gleason_thesis}.

\begin{figure}
  	\graphicspath{ {./} }
   	\includegraphics[width=0.9\textwidth]{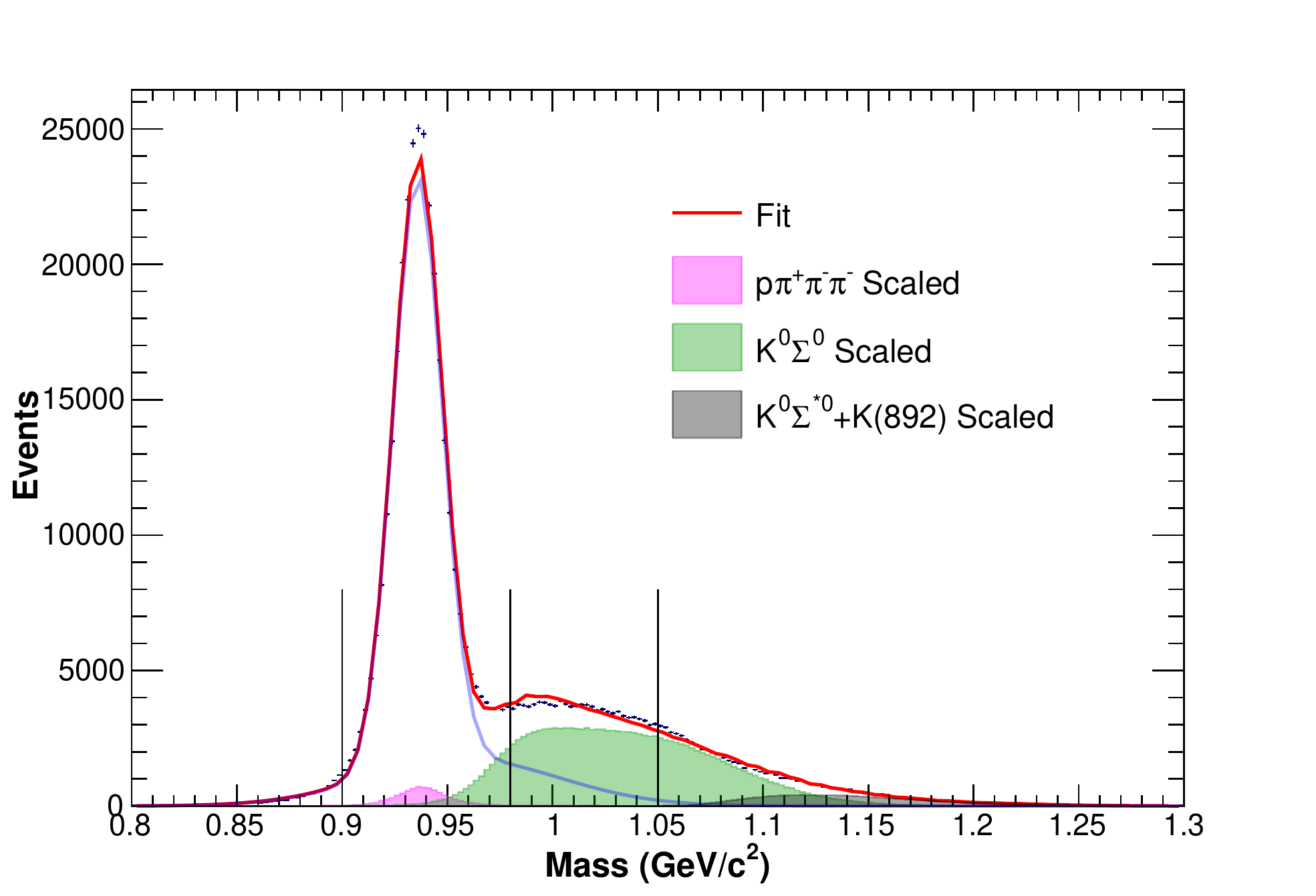} 
	\caption{An example of the fit and scaling to $M_{X}$. The solid red line shows the total fit to the data, the pink histogram is the scaled $p\pi^{-}\pi^{+}\pi^{-}$ distribution, the green histogram is the scaled $K^{0}\Sigma^{0}$ distribution, and the black and violet histograms are the scaled higher-mass channels. The black vertical lines indicate the division of the $M_X$ range into regions for the purpose of background estimates (see text for more details).}
	\label{fig:Mx_data_tot_example}
\end{figure}

The signal-to-background ratios for each background channel were determined by fitting the sum of the simulated background $M_X$ distributions and a double-Gaussian distribution (modeling the signal) to the real-data $M_X$ distribution for each kinematic bin.
The $M_X$ fits yielded the normalization factors needed to scale the simulated backgrounds in order to reproduce the amount of background in real data.
Figure \ref{fig:Mx_data_tot_example} shows an example fit to $M_{X}$.
The solid red line shows the total fit to the data, the pink distribution is the scaled unpolarized $p\pi^{+}\pi^{-}\pi^{-}$ background, the green distribution is the scaled $K^{0}\Sigma^{0}$ background, and the black distribution represents the scaled higher-mass channels.

The three vertical black lines divide $M_{X}$ into two regions: a signal-dominated region, Region 1, which covers the range between the left and the middle black line, 
and a background-dominated region, Region 2, which covers the range between the middle and the right black line.
The background-free observable, $O^{S}$ ($S$ stands for signal), is calculated by using ratios of signal yield to polarized-background yield, $r_{i}^{B}$, and to unpolarized-background yield, $r_{i}^{unpol}$,
\begin{equation}\label{NT_CxSf}
	O^{S}=\frac{r_{1}^{B}O_{2}^{T}-r_{2}^{B}O_{1}^{T}}{r_{1}^{B}-r_{2}^{B}-r_{1}^{B}r_{2}^{unpol}+r_{1}^{unpol}r_{2}^{B}},
\end{equation}	
where $i=1$ or $2$ labels the region \cite{Zachariou_FSI_2016}.

\subsection{Maximum Log-Likelihood Method}\label{sec:ML_Method}
The observables $C_{x}^{T}$, $C_{z}^{T}$, and $P^{T}$ were simultaneously extracted using a maximum likelihood estimator.
For each event $i$, the likelihood function, $L(x,y,z)$, is expressed as
	\begin{equation}
		L^{\pm}_{i}(x,y,z)=1\pm a\cos\theta_{x,i}\pm b\cos\theta_{z,i}+c\cos\theta_{y,i},
	\end{equation}
where $\theta_{x,y,z}$ are the decay angles of the proton in the $\Lambda$ rest frame with respect to the $x$, $y$, and $z$ axes and $+$ or $-$ denotes the helicity of the photon beam.
The best solution for the parameters $a, ~b$, and $c$ is the set that maximizes the log of the likelihood function:
	\begin{equation}
		log~L=\sum_{i=1}^{N^{+}}log(L^{+}_{i})+\sum_{i=1}^{N^{-}}log(L^{-}_{i}),
	\end{equation}
where $N^+$ and $N^-$ denote the number of events obtained with photons of positive and negative helicity, respectively, in the total sample.
The observables are then derived from the parameters $a$, $b$, and $c$ as
	\begin{equation}
		C_{x}^T=\frac{a}{\alpha P_{circ}}, ~C_{z}^T=\frac{b}{\alpha P_{circ}}, ~P^T=\frac{c}{\alpha},
	\end{equation}
where $P_{circ}$ is the degree of circular polarization and $\alpha$ is the self--analyzing power of the $\Lambda$.
In this estimator, it is assumed that the polarization observables do not depend on the detector acceptance.
This assumption, which may lead to a systematic bias in the final value of the observable, was studied extensively with simulated data and realistic polarized cross sections \cite{gleason_thesis}. 
The results show that the detector acceptance has a very weak effect and only on the observable $P$. 
This small bias is included in the systematic uncertainty \cite{gleason_thesis}.

\section{Preliminary Results}
\label{sec:Results}
\subsection{Observables as Functions of $W$}
\begin{figure}[!htb]
  	\centering
  	\graphicspath{ {./} }
   	\includegraphics[width=0.9\textwidth]{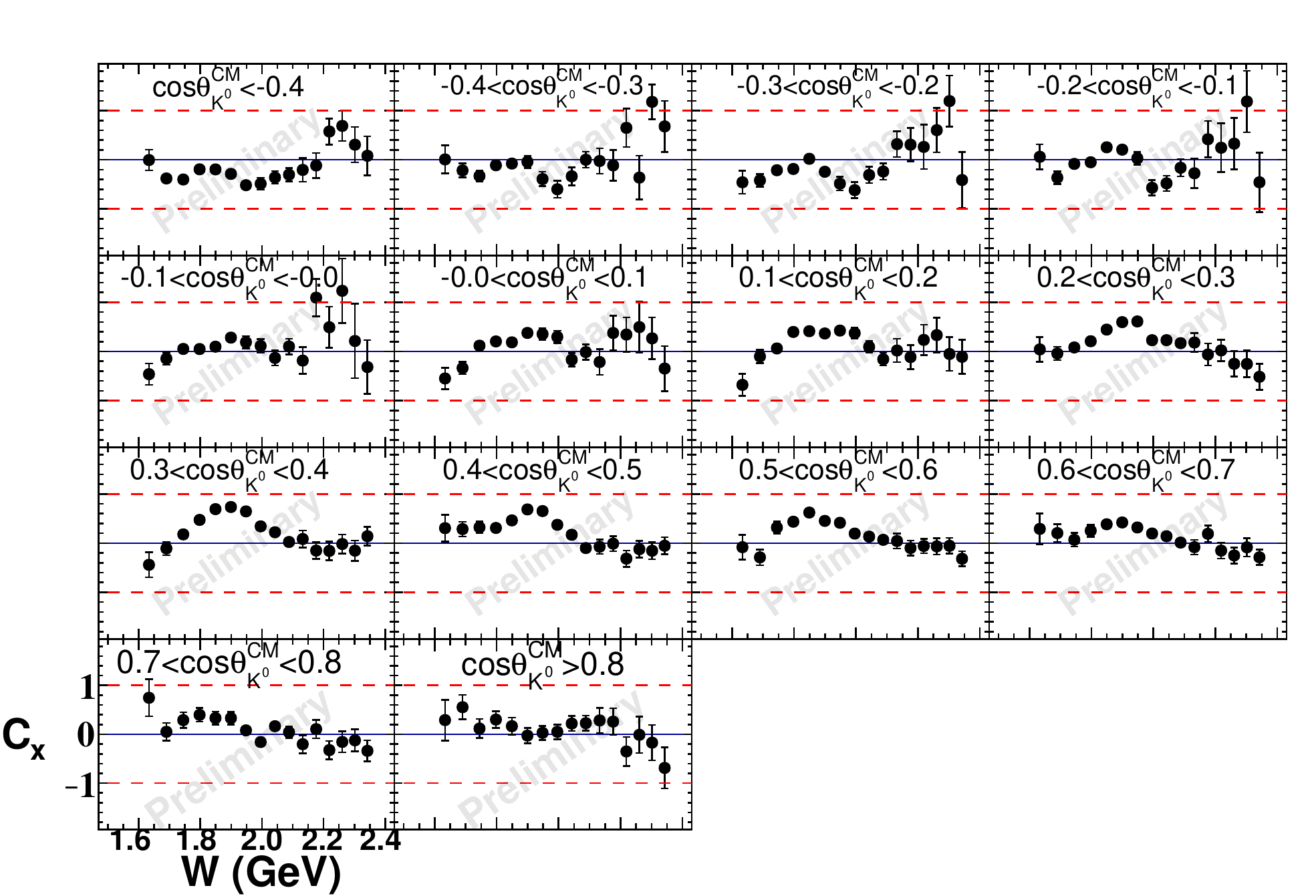} 
	\caption{Preliminary results for $C_{x}$ as a function of $W$ for different $\cos\theta_{K^{0}}^{CM}$ bins. The error bars show statistical uncertainties only.}
	\label{fig:CxW}
\end{figure}

\begin{figure}[!htb]
  	\centering
  	\graphicspath{ {./} }
   	\includegraphics[width=0.9\textwidth]{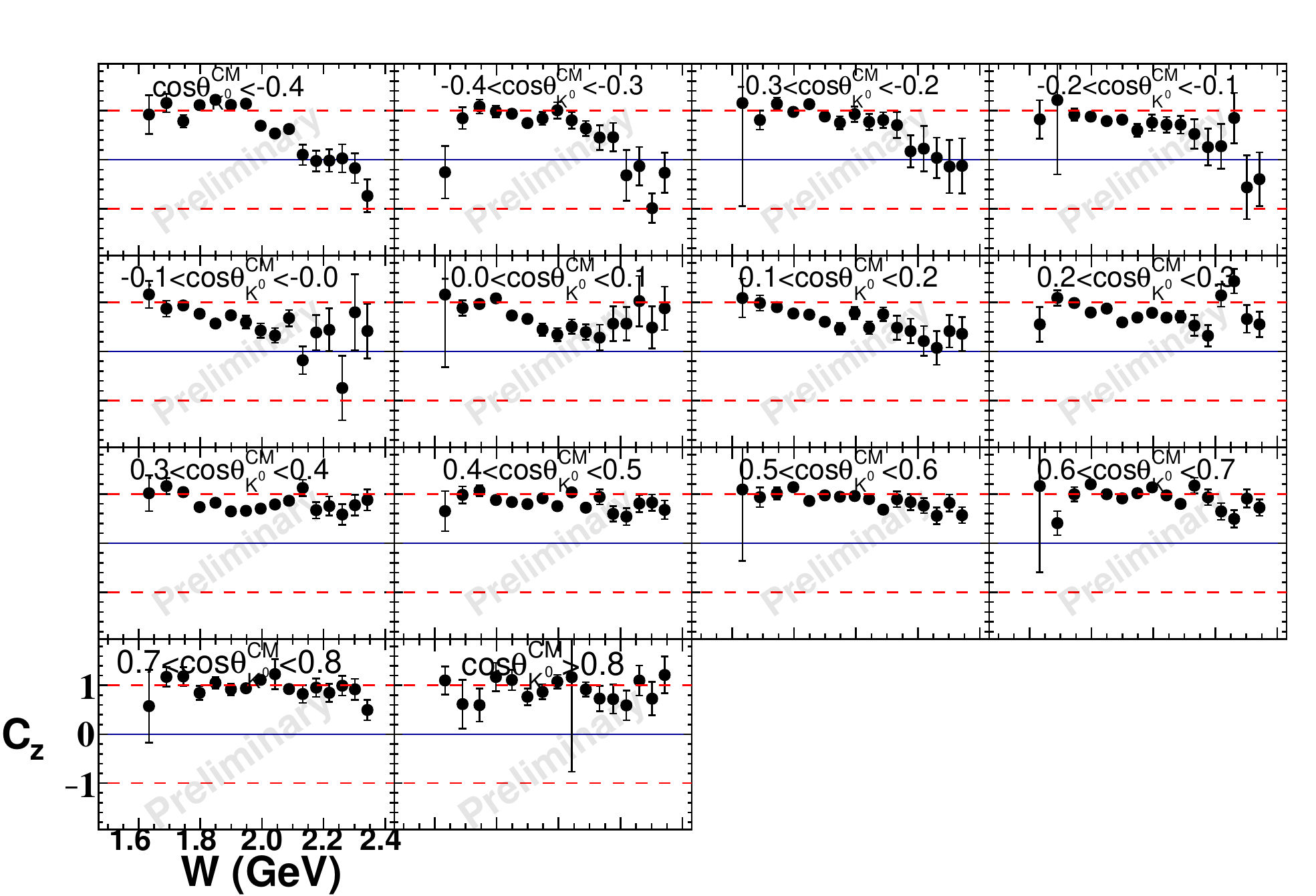} 
	\caption{Preliminary results for $C_{z}$ as a function of $W$ for different $\cos\theta_{K^{0}}^{CM}$ bins. The error bars show statistical uncertainties only.}
	\label{fig:CzW}
\end{figure}

\begin{figure}[!htb]
  	\centering
  	\graphicspath{ {./} }
   	\includegraphics[width=0.9\textwidth]{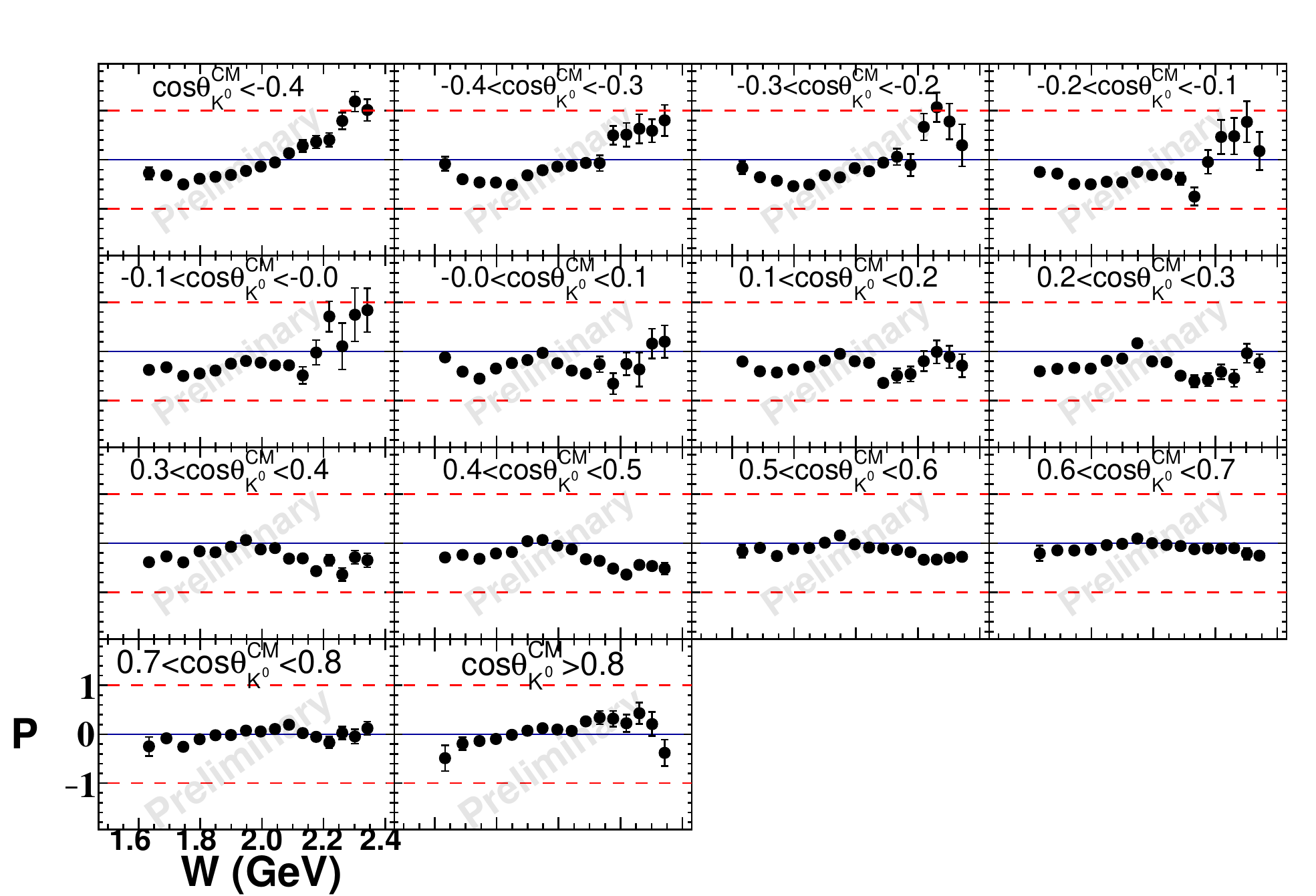} 
	\caption{$P$ as a function of $W=\sqrt{m_{n}^{2}+2m_{n}E_{\gamma}}$ for the 14 different $\cos\theta_{K^{0}}^{CM}$ bins. The error bars show statistical uncertainties only.}
	\label{fig:PW}
\end{figure}

The observables were extracted for 16 $W$ bins and 14 $\cos\theta_{K^{0}}^{CM}$ bins.
Figures \ref{fig:CxW}, \ref{fig:CzW}, and \ref{fig:PW} show preliminary estimates of $C_{x}$, $C_{z}$, and $P$ as a function of $W$ for the 14 $\cos\theta_{K^{0}}^{CM}$ bins.
At backward angles, $C_{x}$ is close to 0.
For $0.0<\cos\theta_{K^{0}}^{CM}<0.8$, $C_{x}$ gradually rises at low energies, peaks around $W\approx1.9$ GeV, then gradually decreases at the highest energies.
Above $\cos\theta_{K^{0}}^{CM}$ of 0.8, $C_{x}$ is fairly constant throughout all $W$ bins.

One interesting, and unpredicted, feature of the experimentally obtained $C_{z}$ in $\gamma p\to K^{+}\Lambda$, was that it was large and close to 1 over all kinematics \cite{Bradford:2006ba}.
This means that the $\Lambda$ was fully polarized along the $z$--axis. 
This feature, when further examined by the Bonn-Gatchina PWA group, showed that the large polarization is due to an intermediate $P_{13}$ $N^{*}$ state, later known as the $N(1900)\frac{3}{2}^{+}$ \cite{Anisovich:2007bq}.
At forward angles (above $\cos\theta_{K^{0}}^{CM}$ of 0.6), the $K^{0}\Lambda$ data show features similar to those of $K^{+}\Lambda$ data, \textit{i.e.}, $C_{z}$ is close to 1 and does not drastically change as a function of $W$.
However, at backward angles, this is not the case throughout all $W$ bins.

Like $C_{x}$, $P$ is generally close to 0 and has some notable features around $W\approx2$ GeV.
At backward angles, $P$ starts out negative at low $W$ then gradually increases to become large and positive at high $W$.
At mid and slightly forward angles, $P$ has a local maximum at $W\approx1.9$ GeV.
For the most forward angles, $P$ is fairly constant and close to 0.

\begin{figure}
  	\centering
  	\graphicspath{ {./} }
   	\includegraphics[width=0.9\textwidth]{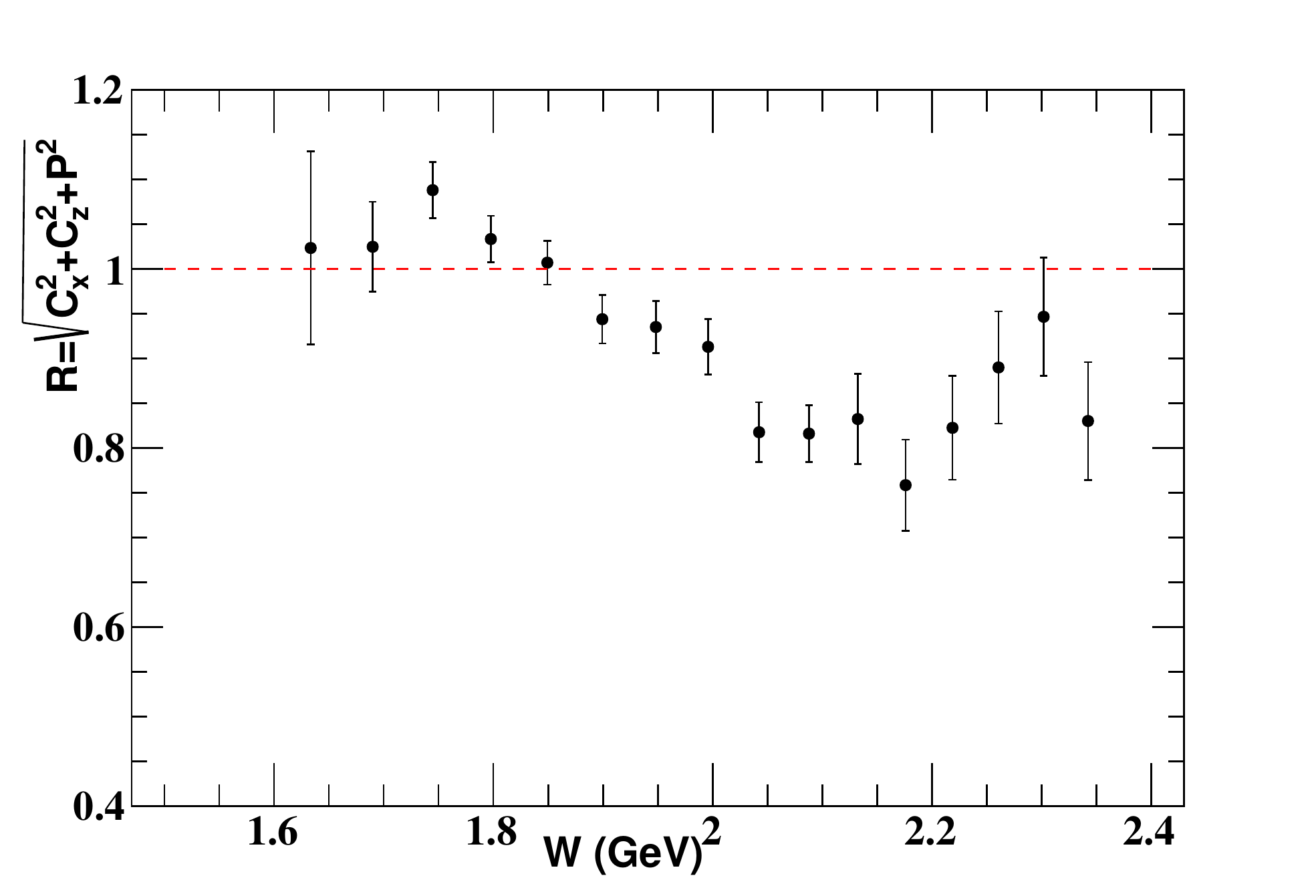} 
	\caption{The quantity $R=\sqrt{C_{x}^{2}+C_{z}^{2}+P^{2}}$ as a function of $W$. Each data point is obtained after averaging over all $\cos\theta_{K^{0}}^{CM}$. The error bars show statistical uncertainties only.}
	\label{fig:RW}
\end{figure}

Another interesting correlation is the energy dependence of the quantity $R=\sqrt{C_{x}^{2}+C_{z}^{2}+P^{2}}$, which can be interpreted as the magnitude of the $\Lambda$ polarization for a 100\% circularly-polarized photon beam.
A value of $R=1$ means that the outgoing $\Lambda$ hyperons are fully polarized.
Figure \ref{fig:RW} shows $R$ averaged over all $\cos\theta_{K^{0}}^{CM}$ for each $W$ bin.
Throughout the lowest $W$ bins, $R$ is consistent with 1.
One $W$ bin has a polarization larger that 1, but the statistical and the systemic uncertainties (not included in this plot) allow for this to happen.
At higher $W$ bins, $R$ drops down to 0.8.
This is different than what was observed in the free-proton data, where $R=1$ for nearly all kinematic bins \cite{Bradford:2006ba}.

Figures \ref{fig:CxW}--\ref{fig:RW} show only the statistical uncertainties of the observables. 
The total systematic uncertainty is bin dependent and is $\leq 0.075$ for $C_{x}$, $\leq 0.120$ for $C_{z}$, and $\leq 0.099$ for $P$. 
The major source of systematic uncertainty for all observables is the background subtraction method, including uncertainties in the simulation, such as detector resolution and background shapes. 
The selection of specific analysis cuts, such as missing-momentum cut, fiducial cuts, and $K^0$ and $\Lambda$ invariant mass cuts, are another source of systematic uncertainty. 
A full treatment of the systematic uncertainties is given in Reference \cite{gleason_thesis}.

\subsection{Dependence on Neutron Momentum}
We are interested in polarization observables for photoproduction off the free neutron, however our data are obtained for a quasi-free production, which
means the neutron is not free but bound with the proton to make the deuteron.
The fact that the neutron has non-zero momentum and is bound can have some effect on the extracted polarization observables. by
The dependence of the observables on missing momentum within the selection range of  $<0.2$ GeV/$c$ can be studied looking at the observables as a function of neutron momentum.
If the observables are constant over the quasi-free selection range (or a subrange), then they should accurately represent what would be seen with a free neutron target.
If the observables are not constant over the quasi-free range, then it may be possible to extrapolate to the ``free" neutron point of $|p_{neutron}|=0$ GeV/$c$ by fitting the functional dependence of 
each observable on $|p_{neutron}|$.

\begin{figure}[!htb]
   \centering
   \graphicspath{ {./} }
   \includegraphics[width=0.8\textwidth]{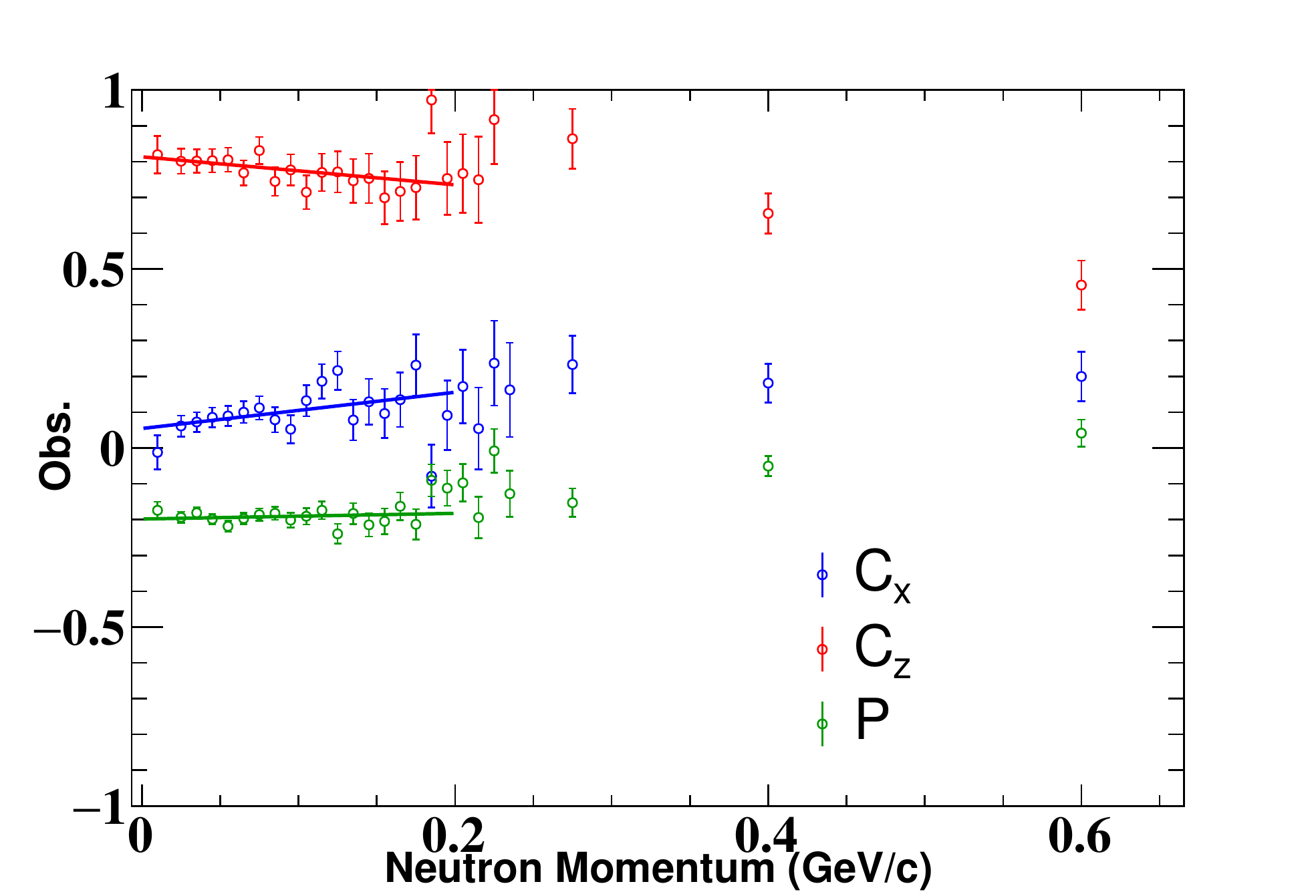} 
   \caption{$C_{x}$ (blue), $C_{z}$ (red), and $P$ (green), integrated over all $W$ and $\cos\theta_{K^{0}}^{CM}$, as a function of neutron momentum. Each observable is fit with a first-order polynomial. The error bars show statistical uncertainties only.} 
   \label{fig:Cx_Cz_P_neutronmom}
\end{figure}

Figure \ref{fig:Cx_Cz_P_neutronmom} shows $C_{x}$ (blue), $C_{z}$ (red), and $P$ (green), integrated over all $W$ and $\cos\theta_{K^{0}}^{CM}$, as a function of neutron momentum.
Each observable is fit with a first-order polynomial.
Our results suggest a weak dependence of $C_x$ and $C_z$ with neutron momentum, whereas 
$P$ seems to be independent of neutron momentum.
Within the uncertainties of the fit parameters, $C_{x}$ has a slope consistent within 2$\sigma$ with 0, $C_{z}$ within 3$\sigma$ with 0, and $P$ within one $\sigma$ with 0.
These results suggest that the cut of 0.2 GeV/$c$ on the missing momentum may not have a large effect overall on the observables. Of course, the neutron-momentum dependence of each observable needs to be
verified for each kinematic bin before any conclusive statement can be made.

\subsection{Legendre Polynomial Fits}
One way to learn more from our results is to perform Legendre polynomial fits to the observables $\mathcal{O}=O\frac{d\sigma}{d\Omega}$ \cite{Wunderlich:2016imj}.
While this is by far not as rigorous as a PWA, it enables one to determine the highest dominant partial wave.
The expansions of $\mathcal{C}_{x},~\mathcal{C}_{z}$, and $\mathcal{P}$ that will be fit here are
\begin{equation}\label{eq:Cx_expansion}
	\mathcal{C}_{x}=\rho\sum_{k=1}^{2L_{max}+1}(a_{L})_{k}^{\mathcal{C}_{x}}P_{k}^{1}(\cos\theta),
\end{equation}
\begin{equation}\label{eq:Cz_expansion}
	\mathcal{C}_{z}=\rho\sum_{k=0}^{2L_{max}+1}(a_{L})_{k}^{\mathcal{C}_{z}}P_{k}^{0}(\cos\theta),
\end{equation}
\begin{equation}\label{eq:Py_expansion}
	\mathcal{P}=\rho\sum_{k=1}^{2L_{max}}(a_{L})_{k}^{\mathcal{P}}P_{k}^{1}(\cos\theta),
\end{equation}
where $\rho=\frac{p}{q}$ is a phase-space factor ($p$ and $q$ denote the center-of-mass momenta of the $\gamma$ and $K^{0}$, respectively), $(a_{L})_{k}^{O}$ are energy-dependent expansion coefficients, and $P_{k}^{i}(\cos\theta)$ are the associated Legendre polynomials without the Condon-Shortley phase.
The highest dominant partial wave component can be found by finding which $L_{max}$ yields $\chi^{2}\to1$.

The fits to our data show that for $W<2$ GeV, the highest dominant partial wave contributing to $\mathcal{C}_{x}$ is of $L_{max}=1$. 
In this mass range, there are the $\star\star~N(1880)\frac{1}{2}^{+}$ and $\star\star\star~N(1900)\frac{3}{2}^{+}$ that are predicted to decay into $K\Lambda$ states. 
It is worth mentioning that the $N(1900)\frac{3}{2}^{+}$ was promoted to a $\star\star\star$ state based on polarization data for $\gamma p\to K^{+}\Lambda$ \cite{Bradford:2006ba}.
Above 2 GeV, $\mathcal{C}_{x}\approx0$ for all bins, so it is hard to determine the dominant contribution in this $W$ range.

$\mathcal{C}_{z}$ has a dominant partial wave of $L_{max}=2$ for nearly all kinematic bins.
There are many states in the range of 1600--2200 GeV: $\star\star\star\star N(1675)\frac{5}{2}^{-},~\star\star\star N(1700)\frac{3}{2}^{-},~\star\star\star N(1875)\frac{3}{2}^{-},~\star\star N(2120)\frac{3}{2}^{-}$.
Two states of particular interest and history for $\gamma n\to K^{0}\Lambda$ are the $\star\star\star N(1875)\frac{3}{2}^{-}$ and $\star\star N(2120)\frac{3}{2}^{-}$ \cite{PDG_2016}.

$\mathcal{P}$ has a dominant partial wave of $L_{max}=1$ for all $W$ bins. 
This is likely due to $P$ being small throughout most kinematic bins.

\section{Summary}
Polarization observables for exclusive photoproduction of strange final states, such as $K\Lambda$, are important to study $N^{*}$ states in the mass range above 1800 MeV/$c^2$.
While good amount of data exists for $K^{+}\Lambda$ photoproduction off the proton, data for $K^{0}\Lambda$ production off the neutron are scarce.
A PWA fit to the differential cross section of $\gamma d\to K^{0}\Lambda(p)$ yielded two solutions that described the data well.
Including polarization observables in this fit could either resolve the ambiguity between the two solutions or provide new results.

This work presents the first-ever estimates of $C_{x}$, $C_{z}$, and $P$ for photoproduction of $K^0\Lambda$ off the bound neutron by measuring the reaction $\vec{\gamma} d\to K^{0}\vec{\Lambda}(p)$.
Our data cover a range of $W$ between 1.65 GeV and 2.35 GeV and a range of $\cos\theta_{K^{0}}^{CM}$ between $-0.9$ and 1.
While the overall features of the kinematical dependences of the observables are very similar to those observed for $K\Lambda$ photoproduction off the proton, there are some notable diferences.
At mid center-of-mass angles, $C_x$ shows a clear peak around $W=1.9$ GeV and is not consistent with 0 at these kinematics. There are kinematics where $C_z$ is not consistent with 1 -- 
at backward center-of-mass angles $C_z$ decreases from 1 to 0 as $W$ increases.
Additionally, $R\neq1$ for $W$ between 2 GeV and 2.2 GeV. 

Eventually, the results of this work will be fit in a PWA. The PWA fits of the Bonn-Gatchina group to the previously published $K^0\Lambda$ cross sections did not show any evidence for new $N^{*}$ states.
While it is too early to discuss whether or not the polarization observables presented here will do so, they should be able to provide additional constraints to current $N^{*}$ amplitudes.


\bibliographystyle{spphys}       
\bibliography{mybibnew.bib}   

\begin{thebibliography}{10}
\providecommand{\url}[1]{{#1}}
\providecommand{\urlprefix}{URL }
\expandafter\ifx\csname urlstyle\endcsname\relax
  \providecommand{\doi}[1]{DOI \discretionary{}{}{}#1}\else
  \providecommand{\doi}{DOI \discretionary{}{}{}\begingroup
  \urlstyle{rm}\Url}\fi

\bibitem{PhysRevD.58.074011}
S.~Capstick, W.~Roberts, Phys. Rev. D \textbf{58}, 074011 (1998)

\bibitem{PhysRevD.94.074040}
R.~Bijker, J.~Ferretti, G.~Galat\`a, H.~Garc\'{\i}a-Tecocoatzi, E.~Santopinto,
  Phys. Rev. D \textbf{94}, 074040 (2016)

\bibitem{Bradford:2006ba}
R.~Bradford, et~al., Phys. Rev. C \textbf{75}, 035205 (2007)

\bibitem{McNabb:2003nf}
J.W.C. McNabb, et~al., Phys. Rev. C \textbf{69}, 042201 (2004)

\bibitem{PhysRevC.88.035209}
H.~Kamano, S.X. Nakamura, T.S.H. Lee, T.~Sato, Phys. Rev. C \textbf{88}, 035209
  (2013)

\bibitem{Compton:2017xkt}
N.~Compton, et~al., Phys. Rev. C \textbf{96}, 065201 (2017)

\bibitem{NadelTuronski2008}
P.~Nadel-Turo{\'{n}}ski, Few Body Syst. p. 227 (2008)

\bibitem{proposal}
P.~Nadel-Turo{\'{n}}ski, et~al., Jefferson Lab Experiment E06-103  (2006)

\bibitem{Sober2000263}
D.I. Sober, et~al., Nucl. Instrum. Methods Phys. Res. \textbf{A 440}, 263
  (2000)

\bibitem{CEBAF2_paper}
C.W. Leemann, D.R. Douglas, G.A. Krafft, Ann. Rev. Nucl. Part. Sc. \textbf{51},
  413 (2001)

\bibitem{Mecking2003513}
B.A. Mecking, et~al., Nucl. Instrum. Methods Phys. Res. \textbf{A 503}, 513
  (2003)

\bibitem{LACOMBE1981139}
M.~Lacombe, B.~Loiseau, R.V. Mau, J.~C\^{o}t\'{e}, P.~Pir\'{e}s,
  R.~de~Tourreil, Phys. Lett. B \textbf{101}, 139 (1981)

\bibitem{Zachariou_FSI_2016}
{N. Zachariou, Y. Ilieva, and T. Cao} (2016).
\newblock CLAS-Analysis 2016-003

\bibitem{wolin_1996}
E.~Wolin.
\newblock {GSIM} user's guide version 1.1 (1996).
\newblock
  \urlprefix\url{https://www.jlab.org/Hall-B/document/gsim/userguide.html}

\bibitem{gleason_thesis}
C.~Gleason, Determination of the hyperon induced polarization and
  polarization--transfer coefficients for quasi-free hyperon photoproduction
  off the bound neutron.
\newblock Ph.D. thesis, University of South Carolina (2017)

\bibitem{Anisovich:2007bq}
A.V. Anisovich, V.~Kleber, E.~Klempt, V.A. Nikonov, A.V. Sarantsev, U.~Thoma,
  Eur. Phys. J. \textbf{A 34}, 243 (2007)

\bibitem{Wunderlich:2016imj}
Y.~Wunderlich, F.~Afzal, A.~Thiel, R.~Beck, Eur. Phys. J. \textbf{A53}(5), 86
  (2017)

\bibitem{PDG_2016}
C.~Patrignani, et~al., Chin. Phys. \textbf{C 40}, 100001 (2016)

\end{thebibliography}

%
%

\end{document}